\documentclass[useAMS,usenatbib,aas_macros]{mn2e}

\usepackage{hhline}
\usepackage{lscape}
\usepackage{graphicx}
\usepackage{colortbl}
\usepackage{amssymb}
\usepackage{amsmath}
\usepackage{natbib}
\usepackage{times}
\usepackage{aas_macros}
\usepackage{threeparttable}

\newcommand{\msun}{~\mathrm{M}_{\odot}}

\usepackage{color}

\def\simpropto{\lower.2ex\hbox{$\; \buildrel \propto \over \sim \;$}}
\def\ltsim{\lower.5ex\hbox{$\; \buildrel < \over \sim \;$}}
\def\gtsim{\lower.5ex\hbox{$\; \buildrel > \over \sim \;$}}

\begin{document}
\title[Link between star forming galaxies and DCBH]{Optimal neighbourhood to nurture giants: a fundamental link between star forming galaxies and direct collapse black holes}
\author[B. Agarwal, et al.]{Bhaskar Agarwal$^{1}$\thanks{E-mail:
bhaskar.agarwal@uni-heidelberg.de}, Fergus Cullen$^{2}$, Sadegh Khochfar$^2$, Daniel Ceverino$^{3,4,5}$, \newauthor{Ralf S. Klessen$^{1,5}$}
\\$^1$Universit{\"a}t Heidelberg, Zentrum f{\"u}r Astronomie, Institut
f{\"u}r Theoretische Astrophysik, Albert-Ueberle-Str. 2, D-69120 Heidelberg \\
$^2$Institute for Astronomy, University of Edinburgh, Royal Observatory, Edinburgh, EH9 3HJ\\
$^{3}$Cosmic Dawn Center (DAWN) \\
$^{4}$Niels Bohr Institute, University of Copenhagen, Lyngbyvej 2, 2100, Copenhagen $\mbox{\normalfont\O}$, Denmark \\
$^5$Universit{\"a}t Heidelberg, Interdiszipli{\"a}res Zentrum f{\"u}r Wissenschaftliches Rechnen, Heidelberg, Germany\\}


\date{00 Jun 2014}
\pagerange{\pageref{firstpage}--\pageref{lastpage}} \pubyear{0000}
\maketitle

\label{firstpage}

\begin{abstract}
Massive $10^{4-5}\msun$ black hole seeds resulting from the \textit{direct} collapse of pristine gas require a metal-free atomic cooling halo with extremely low H$_2$ fraction, allowing the gas to cool isothermally in the presence of atomic hydrogen. In order to achieve this chemo-thermodynamical state, the gas needs to be irradiated by both: Lyman-Werner (LW) photons in the energy range $11.2-13.6$~eV capable of photodissociating H$_2$, and $0.76$~eV photons capable of photodetaching H$^-$. Employing cosmological simulations capable of creating the first galaxies in high resolution, we explore if there exists a subset of galaxies that favour direct collapse black hole (DCBH) formation in their vicinity. We find a fundamental relation between the maximum distance at which a galaxy can cause DCBH formation and its star formation rate (SFR), which automatically folds in the chemo-thermodynamical effects of both H$_2$ photo-dissociation and H$^-$ photo-detachment. This is in contrast to the $\sim 3$ order of magnitude scatter seen in the LW flux parameter computed at the maximum distance, which is synonymous with a scatter in `J$_{crit}$'. Thus, computing the rates and/or the LW flux from a galaxy is no longer necessary to identify neighbouring sites of DCBH formation, as our relation allows one to distinguish regions where DCBH formation could be triggered in the vicinity of a galaxy of a given SFR. 

 \end{abstract}

\begin{keywords}
quasars: general, supermassive black holes -- cosmology: darkages, reionization, firststars -- galaxies: high-redshift
\end{keywords}

\section{Introduction}

The theory of massive black hole (BH) seed formation via direct collapse of pristine gas has undergone drastic revision in the recent years. A direct collapse black hole (DCBH) can form when pristine gas in an atomic cooling halo is rid of its molecular hydrogen (H$_2$), as a result of which atomic hydrogen (H) is the only available coolant \citep{Omukai:2001p128,Oh:2002p836}. The gas can then collapse isothermally at $\sim 8000$~K due to cooling by H, which at a number density of $10^3 \rm \ cm^{-3}$ translates to a Jeans Mass of $10^6\ \msun$ \citep{Regan:2009p776,Latif:2013p2787}. This mass is readily available in atomic cooling haloes and is theorised to withstand fragmentation, thus able to collapse \textit{directly} into a black hole of mass $10^{4-5} \ \msun$. An important step in this process is the destruction of H$_2$. Photons in the energy range 11.2 - 13.6 eV, or the so called Lyman--Werner (LW) band, emanating from a galaxy close to a pristine atomic cooling halo offer a natural way to photo-dissociate H$_2$ \citep{Haiman:2000p87,Omukai:2001p128,Machacek:2001p150,Yoshida:2003p51,OShea:2008p41}.
%

The minimum level of LW radiation field that is able to efficiently dissociate H$_2$ in the pristine atomic cooling halo and facilitate atomic H cooling is thus parameterised in the form of a \textit{critical} specific intensity, J$_{crit}$, normalised to $10^{-21}\rm \ erg\ cm^{-2}\ Hz^{-1}\ sr^{-1}$. Initial studies assumed that the spectrum of first (Population III, Pop III) and second generation (Population II, Pop II) of stars can be represented by a blackbody with a temperature of $10^5$ and $10^4$~K respectively. Thus, a J$_{crit} \sim 30 \ \&\ 1000$ from Pop III and Pop II stellar populations was derived in literature \citep{Omukai:2001p128,Shang:2010p33,2014MNRAS.443.1979L}.
However, the spectrum of the first galaxies is far from a perfect blackbody, and consists of multiple generation of stars with different ages and metallicities. Furthermore, collapse of gas in a pristine atomic cooling haloes does not solely depend on the H$_2$ fraction as it has been shown that  H$^-$ at low densities can catalyse H$_2$ formation, and ignoring H$^-$ destruction can lead to erroneous deductions in the final  chemo-thermodynamical state of the gas \citep[][A16,W17 hereafter]{Agarwal15b,Wolcott2017}. In order to improve upon the J$_{crit}$ parameterisation that hinges solely on the LW band, recent studies employed realistic stellar population synthesis (SPS) schemes to model the spectrum of the first galaxies \citep{Sugimura:2015ut,Agarwal15a} and proposed an alternative way to determine if a pristine atomic cooling halo can undergo DCBH formation (A16; W17). The new scheme replaces a single critical value of J$_{crit}$ by a critical curve in the H$_2$ - H$^-$ destruction rate parameter space, where regions above the curve represent the combination of destruction rates that allow the gas to undergo isothermal collapse, and the values that fall below the curve most likely cause the gas to fragment  and form Pop III stars instead (A16; W17).
Although the curve improves upon the simplistic J$_{crit}$ parameterisation, it suggests that no single J$_{crit}$ exists in nature and can vary from $0.1-10^4$ depending on the shape of the SED. In other words, this presents a major uncertainty for cosmological simulations (or semi-analytical-models) that aim to model the abundance of DCBHs in the early Universe.

This study focuses on extracting galaxies from cosmological simulations, and analysing their self-consistent spectral energy distributions (SEDs) in order to determine their position on the H$_2$-H$^-$ destruction rate parameter space. This is the first time that realistic star formation histories (SFH) have been used to determine the SED of a galaxy in order to explore conditions conducive for DCBH, as all previous studies have used idealised SFH in order to do the same \citep{Agarwal15a,Sugimura:2015ut}.
This exercise reveals what subset of these galaxies are most likely to give rise to conditions conducive for DCBH in their local neighbourhood. Furthermore, we are looking for physical parameters (or correlations) that best characterise the SEDs of these galaxies and indirectly capture the effects of the critical curve, thus offering a direct way to identify galaxies that can cause DCBH formation in their vicinity.

\begin{figure}
\includegraphics[width=0.5\textwidth,trim={0.5cm 0.5cm 0.5cm 0.5cm},clip]{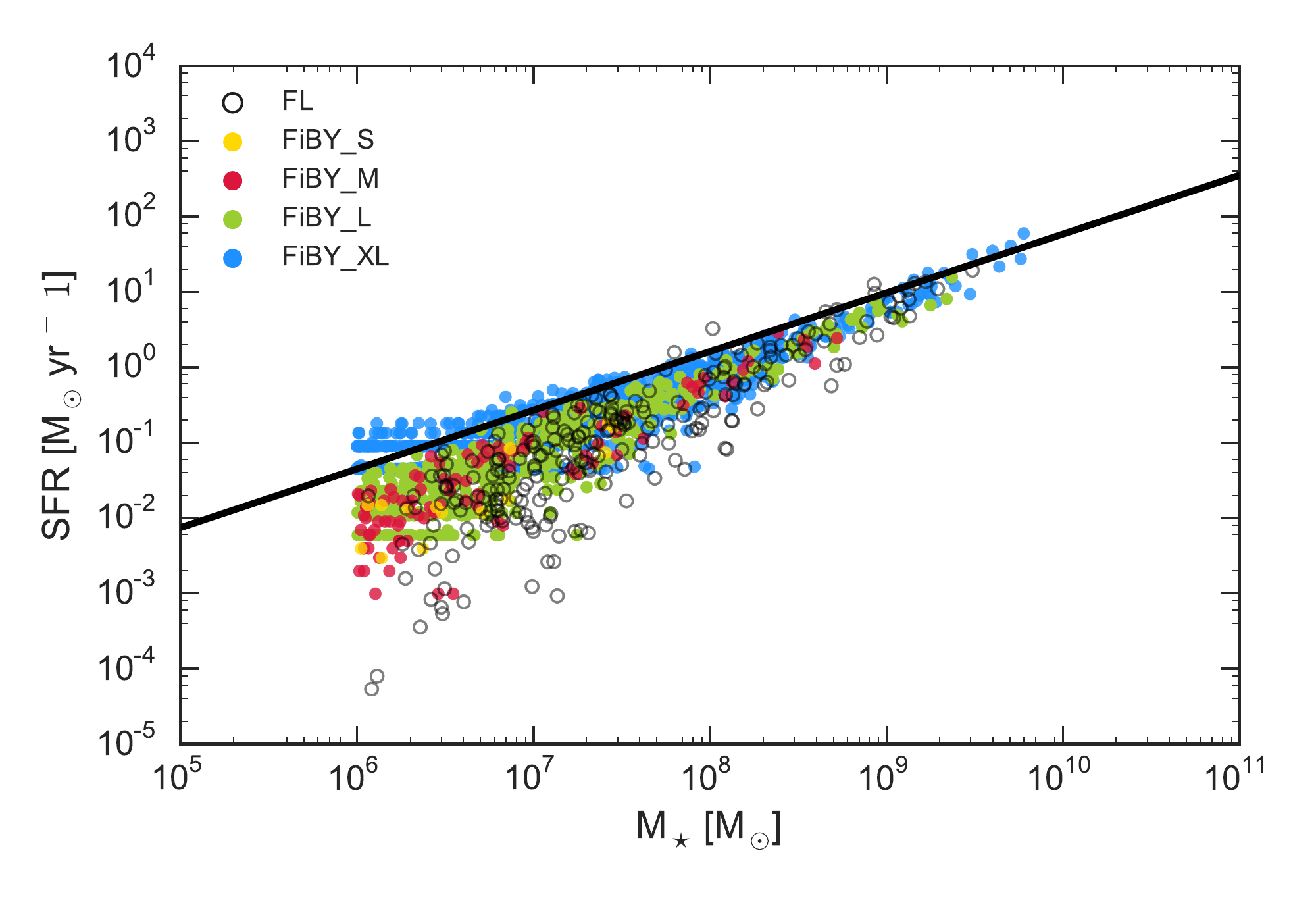}
\caption{The star formation rate and stellar mass main sequence plane for galaxies at $z\sim6$ in our sample. The solid line is the expected MS relation computed at $z=6$ \citep{2014ApJS..214...15S}. 
}
\label{fig.MS}
\end{figure}


\begin{figure*}
\includegraphics[width=0.99\textwidth,trim={2cm 0.5cm 3cm 12.5cm},clip]{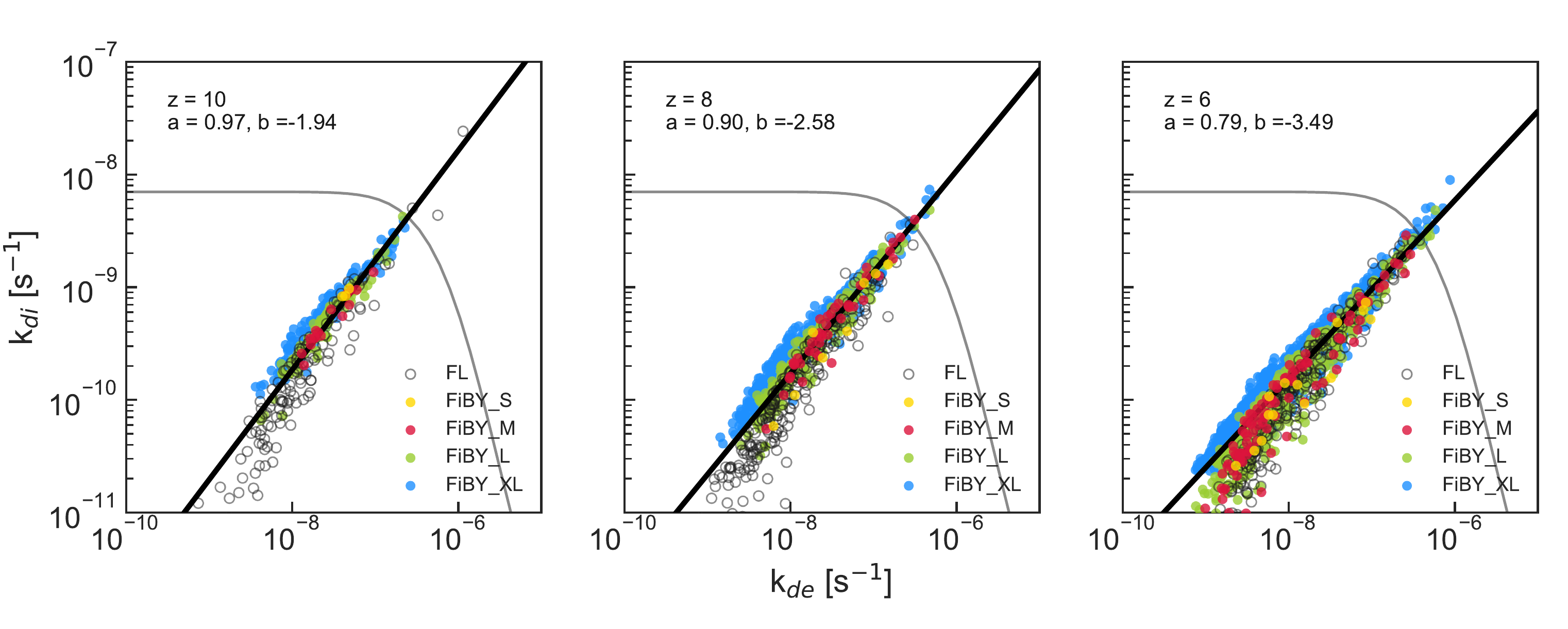}
\caption{Galaxies in the H$_{2}$ photodissociation (k$_{di}$) and H$^{-}$ photodetachment (k$_{de}$) rate plane. The rates (and thus the J$_{LW}$) is computed at half the virial radius of a galaxy's DM halo. This is an arbitrary choice and bears no reflection on the physical extent of the galaxy. In grey we plot the critical curve for DCBH formation (Eq.~\ref{eq.ratecurve}), and in solid black we show the fit to the sample at that redshift. The fit can be parameterised as ${\rm log_{{10}}(k}_{di}) = {\rm alog_{{10}}(k}_{de}) \rm + b$, where a and b are listed under the redshift in each panel. The colour coding is the same as used in Fig. 1.}
\label{fig.dedi}
\end{figure*}

\begin{figure}
\includegraphics[width=0.475\textwidth,trim={0.5cm 0.5cm 0.5cm 0.5cm},clip]{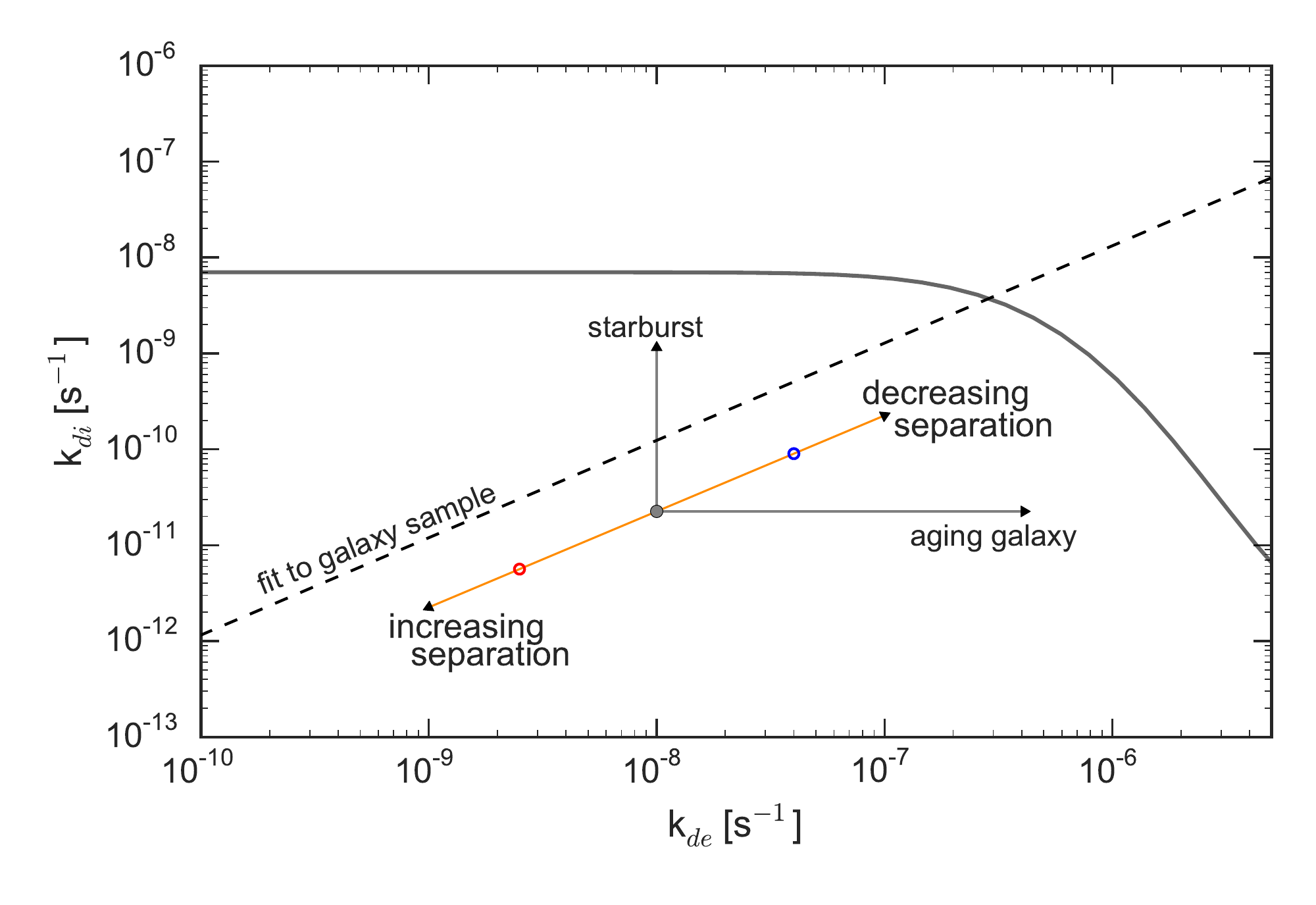}
\caption{Galaxy tracks in the rate plane. At any given redshift, a galaxy (gray dot) can move along the tracks (marked by arrows) representative of a physical process. The dashed line is the fit to our distribution (as seen in Fig.~\ref{fig.dedi}) and in grey we show the critical curve. The orange track has a slope $=1$ (and thus almost parallel to the fit) is obtained by varying the distance in the J$_{LW}$ computation, for example, red (blue) open circle is obtained if the distance between the galaxy and the atomic cooling halo is doubled (halved).}
\label{fig.scaling}
\end{figure}

\begin{figure}
\includegraphics[width=0.475\textwidth,trim={0.5cm 0.5cm 0.5cm 0.5cm},clip]{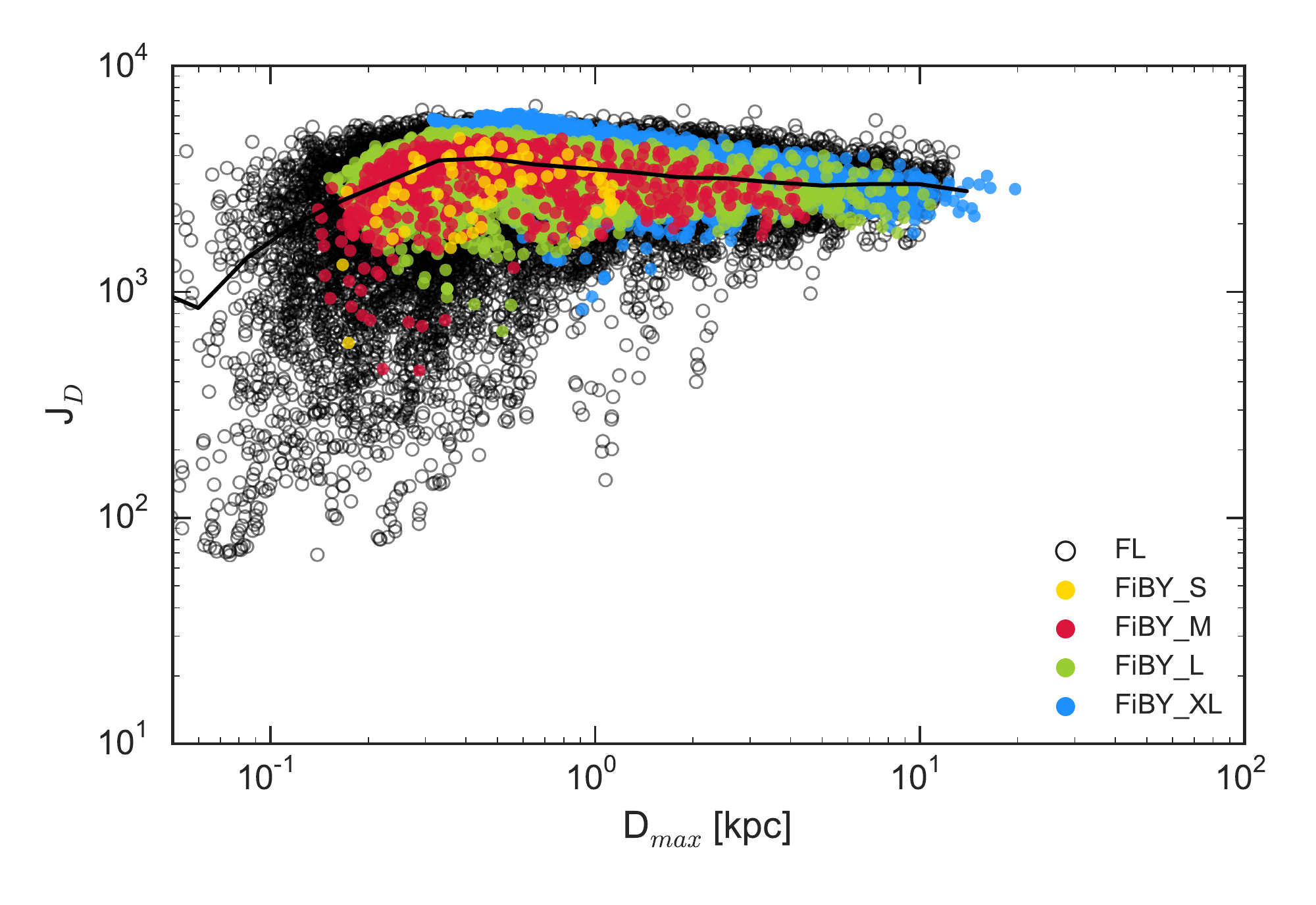}
\caption{The J$_{LW}({\rm D}_{max})$ or J$_D$ vs. D$_{max}$, {for all galaxies at all redshifts in the respective simulation. Or in other words, on the x-axis we show the maximum distance at which  galaxy can cause DCBH formation in an atomic cooling halo, and on the y-axis the LW specific intensity at that distance.} The black solid line denotes the {mean} of the distribution.}
\label{fig.d_Jc}
\end{figure}

\begin{figure*}
\includegraphics[width=0.99\textwidth,trim={2cm 0.5cm 3cm 12.5cm},clip]{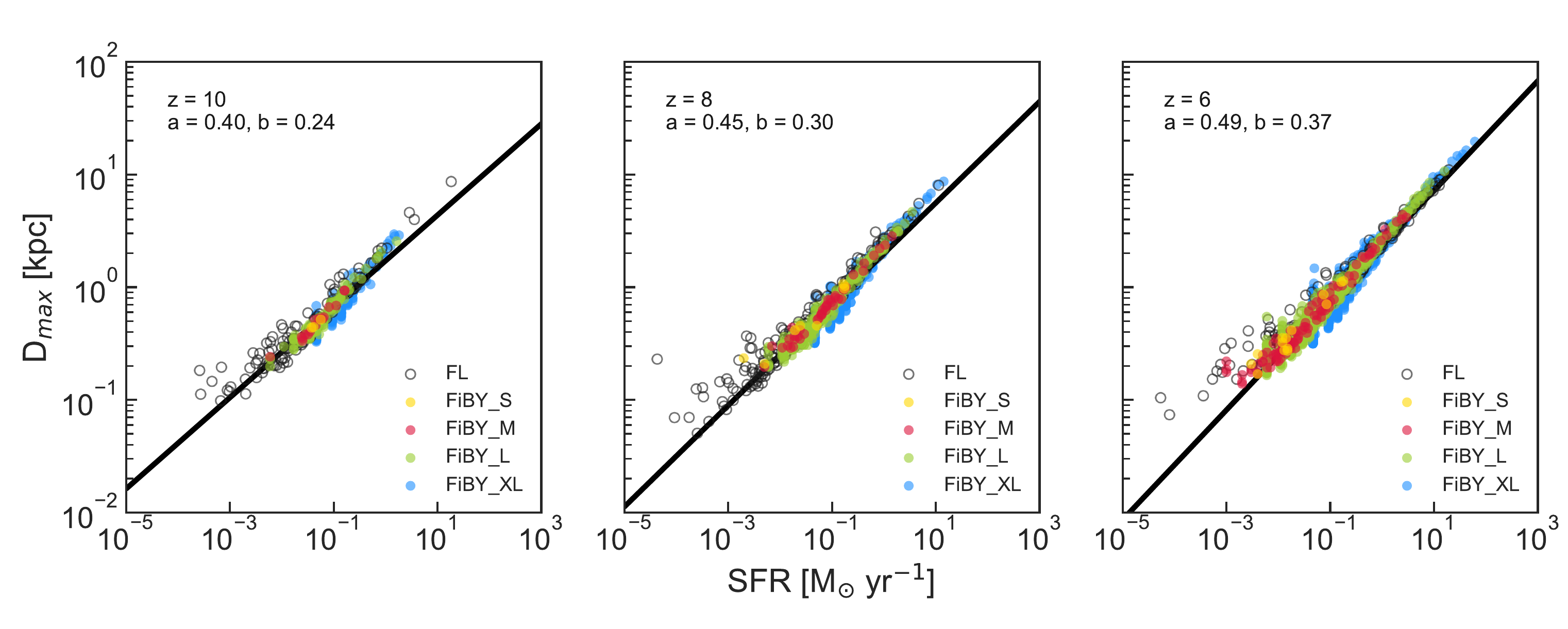}
\caption{The relation between the SFR of a galaxy and D$_{max}$. The solid line denotes the linear fit obtained for the galaxy sample at the corresponding redshift. The fit at each redshift can be parameterised as ${\rm log_{10}(D}_{max}) = \rm alog_{{10}}(SFR) + b$, where a and b are listed under the redshift in each panel. The colour coding is the same as the one used in the previous plots.}
\label{fig.sfr_d}
\end{figure*}


\section{Methodology}
\label{methodology}
In order to understand how the radiative conditions, pertaining to LW feedback, can be best characterised from the first galaxies in the Universe, we chose two mutually independent simulation suites that are able to produce galaxies at $z\geq6$ in the stellar mass range $\sim 10^6 - 10^{10} \msun$. 

\subsection{Simulation: overview}

The first one is First Billion Year (FiBY) simulation suite, run using a modified version of {\sc gadget} \citep[used in the Overwhelmingly Large Simulation Project]{Schaye:2010p2481} employing a smoothed-particle-hydrodynamic approach, in boxes with a side length of 4, 8, 16 and 32 cMpc, which we will refer to as FiBY\_S, FiBY\_M, FiBY\_L, and FiBY\_XL from this point onwards. The simulations take into account a wide range of sub-grid physical processes including relevant molecular cooling processes and Pop II and Pop III star formation. They are able to simultaneously reproduce the SFR and stellar mass function, and the physical trends in the evolution of the metal content of galaxies. The FiBY suite has been employed to corroborate cosmic reionisation by low-mass galaxies \citep{Paardekooper:2013p2455,JP2015}, the environment of DCBHs \citep{Agarwal14}, the impact of stellar radiation on Pop III and Pop II stars \citep{Johnson:2013p2049}, constraining the dust attenuation law of star-forming galaxies at high redshift \citep{2017MNRAS.470.3006C}, the $\gamma$-ray burst rate \citep{Jonny12} and the effect of feedback on dark matter density profiles \citep{Andrew14}.

Second, the FirstLight (FL) database is a complete mass-selected sample of central galaxies simulated in 290 cosmological, zoom-in simulations performed with the adaptive mesh refinement (AMR) ART code \citep{1997ApJS..111...73K,2003ApJ...590L...1K,2009ApJ...695..292C}.  Besides gravity and hydrodynamics, the code incorporates many of the astrophysical processes relevant for galaxy formation. These processes, representing subgrid physics, include gas cooling due to atomic hydrogen and helium, metal and molecular hydrogen cooling, photoionization heating by a constant cosmological UV background with partial self-shielding, star formation and feedback (thermal, kinetic \& radiative), as described in \cite{2017MNRAS.470.2791C}. The FirstLight suite has been also employed to unveil the origin of the mean and scatter of  the main sequence of star-forming galaxies during cosmic dawn \citep{2018MNRAS.tmp.2027C}.

We summarise the simulation parameters, for both the suites, in Table.~\ref{tab.sim}, and plot the stellar mass - star formation rate for all the galaxies at $z\sim6$ in Fig.~\ref{fig.MS}. The solid line is meant to denote the \textit{main sequence} (MS) of galaxy evolution, computed at $z=6$ \citep{2014ApJS..214...15S}.

\subsection{SEDs: overview}

\subsubsection{FiBY}
To generate the galaxy SEDs we followed the method outlined in \citet{2017MNRAS.470.3006C}.
Briefly, for each galaxy, the star particles associated with it were assigned an instantaneous starburst SED based on the their individual ages and metallicities. 
The final galaxy SED was then constructed by summing the SED of each individual star particle.
For the SPS models we used the `Binary Population and Spectra Synthesis' version 2.1 \citep[BPASSv2.1;][]{bpassv2.1}. The fiducial model in this study, BPASSv2.1-300bin, includes binary SEDs with an IMF index of $-1.3$ between $0.1 - 0.5$ M$_{\odot}$ and $-2.35$ between 0.5 - 100 M$_{\odot}$. The models span the following metallicities: $Z_{\star}=10^{-5},10^{-4},0.001,0.002,0.003,0.004,0.006,0.008,0.01,0.014$ $,0.02,0.03,0.04$.
Finally, the contribution from nebular continuum is included using \textsc{cloudy} \citep{ferland2017} as described in \citet{2017MNRAS.470.3006C} assuming a maximal nebular contribution (i.e. an escape fraction $f_{esc}=0$).

\subsubsection{First Light}
The SEDs of the FL galaxies \citep{2019MNRAS.484.1366C} were also constructed using BPASSv2.1 {(nebular emission to be included in the upcoming FL studies)}, with a similar IMF to the FiBY suite of simulations. For the FL suite we used a grid of 13 values of metallicities, from $Z=10^{-5}$ to 0.04, and 40 logarithmic bins in ages, between 1 Myr and 100 Gyr. 

\subsection{Computing the H$_2$ and H$^-$ destruction rates}

The H$_{2}$ photodissociation and H$^{-}$ photodetachment rates can be written as
\begin{eqnarray}
&\rm k_{di} = \kappa_{di}\beta J_{LW} \label{eq.h2}  \\
&\rm k_{de} = \kappa_{de}\alpha J_{LW} \label{eq.hm} 
\label{eqs.dedi}
\end{eqnarray}
where $\kappa_{\rm de} \approx 10^{-10} \: {\rm s^{-1}}$ and $\kappa_{\rm di} \approx 10^{-12} \: {\rm s^{-1}}$. 
To determine the H$^{-}$ photodetachment and H$_{2}$ photodissociation rate coefficients, we therefore need to specify three dimensionless parameters: $\alpha$, $\beta$, and J$_{LW}$. The J$_{LW}$ is a dimensionless parameter defined as the specific intensity at 13.6 eV (or in the LW band), normalised to 10$^{-21}$ erg$^{-1}$s$^{-1}$cm$^2$Hz$^{-1}$sr$^{-1}$.
The dimensionless rate parameters $\alpha$ and $\beta$ depend on the spectral shape of the incident radiation field and we refer the reader to \cite{Agarwal15a} \& \citet{Omukai:2001p128} for more details. 

\subsection{The critical curve}

Recently it became clear that instead of an ideal blackbody spectrum, if SEDs from stellar population synthesis models are used to compute the chemo-thermodynamical state of the collapsing gas in the pristine atomic cooling halo, the value of J$_{crit}$ can vary over 3 orders of magnitude \citep{Agarwal15b,Agarwal_binary}. This can be attributed to the shape of the SED \citep{Agarwal15a,Sugimura:2014p3946} because the first galaxies in the Universe are not perfect blackbodies with the same surface temperature for a given stellar type, but rather, have unique spectral shapes depending on their SFH. 

Thus it was proposed that instead of using a singular J$_{crit}$, a critical curve in the H$_2$-H$^-$ rate plane be used to demarcate the regions of DCBH formation from those of fragmentation (possibly into Pop III stars). This critical curve does not depend on the shape of the SED of the irradiating galaxy and for any given SED shape, allows us to determine the chemo-thermodynamical state of a pristine atomic cooling halo exposed to LW radiation (A16, W17).

In this study, we use the critical curve proposed by A16

\begin{equation}
\rm k_{di} = 10^{Aexp\left(\frac{-p^2}{2}\right) + D}\  ({\ \rm s^{-1}}),
\label{eq.ratecurve}
\end{equation}
where $p=\frac{\log_{10}(\rm k_{de}s) - B}{C}$ and $A = -3.864,\ B = -4.763,\ C = 0.773$, and $D = -8.154$.

Haloes with reaction rates that lie above the curve can undergo DCBH formation, while the rates below the curve lead to fragmentation.

\begin{table}
\centering
\caption{Summary of simulations considered in this work. Only galaxies till $z\geq6$ are considered in this study. The first column lists the box length, $box\_l$, while the second and the third columns list the dark matter particle mass and gas particle mass, $m_{DM}\  and\ m_{gas}$ respectively. The last column in this table lists the number of galaxies at $z\sim6$, $n_{gal}^6$, but note that the progenitors of these galaxies $z\sim6$ are also used for our analysis.}
\label{tab.sim}
\begin{tabular}{c || c | c | c | c  }
 Sim.& $box_l$ & $m_{DM}$ & $m_{gas}$ & $n_{gal}^6$\\
& (Mpc) & ($\msun$) & ($\msun$) &  \\
 \hline
 \hline
FiBY\_S & 4 & 6181  & 1271  & 14\\
FiBY\_M & 8 &  $5 \times 10^4$ & 10171  & 98\\
FiBY\_L & 16 & $4 \times 10^5$ & $8.1  \times 10^4$ & 479\\
FiBY\_XL & 32 & $3.2 \times 10^6$ & $6.5  \times 10^5$ & 1052\\
FL & 14 - 28 & $10^4$ & -  &  250 
\end{tabular}
\end{table}
\section{Results and Discussion}
\label{results}


In order to understand what galaxies in our database are most likely to cause DCBH formation in their vicinity, we need to select a point external to the galaxy where the reaction rates (or in other words, the specific intensity J$_{LW}$) will be computed, i.e. the location of the hypothetical pristine atomic cooling halo. We select a point unique to each galaxy at half the virial radius of its own dark-matter halo. This is done purely to understand the distribution of galaxies in the k$_{de}$-k$_{di}$ plane and is not reflective of the galaxy's physical dimension, or the distribution of pristine atomic cooling haloes around it. In Fig.~\ref{fig.dedi} we show the results for galaxies considered in this work, plotted on the H$_2$-H$^-$ rate plane. The galaxies populate a narrow diagonal region in the rate parameter space, and the solid black line represents a fit to the entire sample at a given redshift. The fit can be parameterised as ${\rm log_{{10}}(k}_{di}) = {\rm alog_{{10}}(k}_{de}) \rm + b$, where a and b are listed under the redshift in each panel of Fig.~\ref{fig.dedi}. The slope of the fit shows a marginal deviation from unity, and only a slight evolution in the normalisation of the fit is seen with decreasing redshift. 

The position of the galaxies with respect to the critical curve depends on the SED of the galaxy and the distance at which the flux is computed. This peculiar distribution of galaxies in the H$_2$-H$^-$ destruction rate plane is a consequence of the region they occupy in the stellar mass - star formation rate plane, as both FiBY and FirstLight reproduce galaxies that tend to follow the MS of galaxy formation expected at $z>6$ (see Fig.~\ref{fig.MS}).

\subsection{Galaxies causing DCBH formation in their vicinity}

{A galaxy can occupy any point in the k$_{de}$-k$_{di}$ rate plane, and can move in any of the directions depending on the physical process it undergoes. For example, a starburst  would result in more LW photons, an aging stellar population would give rise to more 1eV photons, a change in the separation between the irradiating galaxy and the point where DCBH formation conditions are computed would change the LW flux. Thus, modifying the physical conditions could lead to the atomic cooling halo experiencing photodestruction rates that intersect the critical curve for DCBH formation, as shown in Fig.~\ref{fig.scaling}.}

We carried out this exercise by computing the factor by which the separation needs to be changed such that H$_2$ and H$^-$ photodestruction rates from the galaxy lie exactly on the critical curve for DCBH formation. Thus, as illustrated in Fig.~\ref{fig.scaling} we always move the galaxy on a track (orange line) {almost\footnote{The orange track would have a slope exactly $=1$. However, for the sake of simplicity, we make the 'almost parallel to the fit' argument due to the fact that the fit, dashed line in Fig.~\ref{fig.scaling}, has a slope $\sim1$.}} parallel to the fit (dashed line) at that redshift.  This factor can then translated to a distance, D$_{max}$, which is the maximum separation from a galaxy at which DCBH formation can ensue in a pristine atomic cooling halo. In other words, for every galaxy we can define a maximum separation within which a pristine atomic cooling halo must be situated such that the atomic cooling halo has the correct H$_2$ and H$^-$ photodestruction rates (see Eqs.~\ref{eqs.dedi}) for DCBH formation.

Since the SED of the galaxy is determined by its SFH we can not vary the rate parameters for a galaxy as they are self-consistently derived from the SED. Determining D$_{max}$ allows us to compute the LW specific intensity produced by the galaxy at that distance, i.e. ${\rm J}_{LW}({\rm D}_{max}) = {\rm J}_D$. In other words, J$_{D}$ is the value of the LW specific intensity incident on an atomic cooling halo located at a separation equal to D$_{max}$ from the galaxy. 
The distribution of J$_D$ vs D$_{max}$ is shown in Fig.~\ref{fig.d_Jc}, where the black line denotes the mean of J$_D$ at a given D$_{max}$. Overall, a 2 order of magnitude spread is seen in J$_D$, where at D$_{max} < 1$~kpc a large scatter is seen in J$_D$, while at D$_{max}>1$~kpc, J$_D$ seems to have a mean value of $\sim3000$, albeit with scatter of 1 order of magnitude. This is a result of the scatter seen for low mass galaxies and their SFR, while for more evolved galaxies the stellar masses and SFR exhibit a rather coeval trend (see Fig.~\ref{fig.MS} \& \ref{fig.d_Jc}). The points in Fig.~\ref{fig.d_Jc} at D$_{max} < 1$~kpc are thus the low luminosity galaxies produced in the simulations. The spread in J$_D$ is not as extreme as A16 reported earlier. This is because A16 considered a large parameter space for the stellar mass, age and SFR of galaxies which was not motivated by simulations but chosen in order to maximise the variation in the shape of the SEDs.

The J$_D$ derived here differs from the conventionally derived J$_{crit}$, as this J$_D$ is unique for each galaxy and is the actual J$_{LW}$ produced by a galaxy at D$_{max}$. To reiterate, the spread in J$_D$ is a result of the dependence of the reaction rates on the shape of the source spectrum, where every galaxy's individual SFH is responsible for shaping its SED and thus the rates derived from it. This must not be confused with the fundamental nature of the critical curve in the H$_2$ - H$^-$ rate plane used to demarcate regions of DCBH formation and Pop III star formation, which does not depend on the source spectrum in any way. 

\subsection{Eliminating the need for J$_{crit}$}

The spread in J$_{D}$ for our galaxy sample derived from large scale simulations further establishes that using a singular LW specific intensity for all galaxies without considering the shape of their SED, as has been done in literature thus far, is an erroneous way to identify sites of DCBH formation. Instead, one should use the critical curve in the H$_2$-H$^-$ rate plane as it is indicative of the chemo-thermodynamical state of the collapsing gas in a pristine atomic cooling halo and thus independent of the spectral shape of the irradiating source. This said, computing the rates experienced by the atomic cooling halo due to the radiation emanating from a nearby galaxy requires the knowledge of the galaxy's SED. Here, we propose an alternative solution which takes into account the entire rate-space calculation and shape of the source SED at the same time. We find that D$_{max}$ correlates well with the SFR of the galaxy, as shown in Fig.~\ref{fig.sfr_d}. We did this exercise at each simulation snapshot, but only a few redshifts are plotted in the figure, where the solid black line is a fit to the data expressed by $\rm log_{10}(\frac{D_{\mathit{max}}}{1 kpc}) = alog_{10}(\frac{SFR}{\msun\rm\ yr^{-1}}) + b$, where a and b are listed below the redshift in each panel. This constitutes a fundamental and easy to use relation between a galaxy's current SFR and the maximum distance (or the radius of the sphere of influence) within which it can cause DCBH formation in a pristine atomic cooling halo. The relation conveniently captures the galaxy's location on the H$_2$-H$^-$ rate plane, and is a natural consequence of the SFH of the galaxy. Our result deems J$_{crit}$ redundant, and presents an elegant way to hunt DCBHs in both theoretical models of galaxy formation, as well as observations.

 


\subsection{Multiple galaxies aiding the formation of DCBHs}

So far in the study, we have concentrated on the radiative field emanating from a single galactic source. However, in reality, the LW radiative field builds up from multiple galaxies in the local neighbourhood \citep{Agarwal12,Agarwal14,Habouzit16hSAM} of the pristine atomic cooling halo. Using the results presented above, we will now derive an expression that allows us to capture the destruction rates from multiple galaxies.

Let us start by expressing J$_{LW}$ as a function of the SFR of the galaxy,
\begin{equation}
\rm J_{\mathit{LW}} = \Theta \frac{SFR}{D^2}\ .
\end{equation}
where SFR is the star formation rate of the galaxy, $\Theta$ is a constant relating the LW-photon production to the star formation rate of a galaxy, and D is the distance from the galaxy. Here, we work under the assumption that  $\Theta$ is qualitatively the same for all galaxies to first order independent of the SED, as is the case in our simulations.

As shown earlier, there exists a maximum distance, D$_{max}$, within which a galaxy of a given SFR can induce DCBH formation, or
\begin{equation}
\rm J_{LW}(D_{\mathit{max}})=J_{\mathit{D}} = \Theta \frac{{SFR}}{D_{\mathit{max}}^2}\ .
\end{equation}

We can then write the following condition for the total specific intensity emanating from an ensemble of $\rm n$ galaxies that have a distance $\rm D$ from a pristine halo,
\begin{equation}
\rm J_{\mathit{tot}} =  \Theta \sum_{\mathit{i=0}}^n\frac{ SFR_\mathit{i}}{D_\mathit{i}^2}\ .
\end{equation}

If we assume the average value of J$_{\mathit{LW}}$ to allow for DCBH formation from all galaxies is $\langle J_\mathit{D}\rangle$,  and expressing the distance from each galaxy as d$_i=$f$_i$D$_{\mathit{max,i}}$ from that galaxy, then 
\begin{equation}
\rm J_{\mathit{tot}} =  \Theta \sum_{i=0}^n\frac{SFR_\mathit{i}}{f_\mathit{i}^2D_{\mathit{max,i}}^2} \sim \langle J_\mathit{D}\rangle \sum_{\mathit{i=0}}^n\frac{1}{f_\mathit{i}^2}\ .
\end{equation}
Thus, in order for the ensemble of galaxies to allow DCBH formation in their vicinity, we need the ratio $\rm \frac{J_\mathit{tot}}{\langle J_\mathit{D}\rangle} \geq1$, or
\begin{equation}
\sum_{i=0}^n\frac{1}{\rm f_\mathit{i}^2} \geq1 \equiv \sum_{\mathit{i=0}}^n\frac{\rm D_{\mathit{max,i}}^2}{\rm D_\mathit{i}^2} \geq1\ .
\label{eq.multiple}
\end{equation}
where D$_{max,i}$ for each galaxy can be determined using its SFR and our fit presented in Fig.~\ref{fig.sfr_d}. The aim of the study is to present a more physically consistent and elegant way to understand DCBH formation in the presence of LW radiation, than has been done in the past using a single J$_{crit}$ parameter. Although, the methodology outlined for the multiplicity argument initially employs a similar ansatz, the result itself is a physically consistent expression that does not require knowledge or computation of a critical specific intensity.
\section{Conclusion}

Recent work has emphasised the need for rejecting the J$_{crit}$ parameter, and instead employing the critical curve in the H$_2$-H$^-$ photodestruction rate phase-space to identify haloes that can undergo DCBH formation. In this study, we explored if there is a more elegant way to identify sites of DCBH formation without the need to explicitly compute the radiation flux or the rates.

We analysed the SEDs of over $30000$ galaxies from cosmological SPH and AMR simulations and placed them on the H$_2$-H$^-$ photodestruction rate phase-space to understand if there exists a subset of galaxies most likely to induce DCBH formation in their vicinity. {Recent studies have shown that a single J$_{crit}$ parameter can not fully capture the chemo-thermodynamical state of the collapsing gas and thus, is not the appropriate physical parameter to distinguish DCBH sites. While a critical-curve in the H$_2$ photodissociation and H$^-$ photodetachment rate space solves this problem (A16, W17), employing the curve is challenging as it requires the knowledge of the SED of the irradiating galaxy and chemical rates. Our study is designed to solve this problem by hunting for physical parameters that are able to capture the physical effects of the critical-curve, without the knowledge of the SED or the chemical rates.} We derived a fundamental relation (at a given redshift) between the SFR of a galaxy and the maximum distance at which it can induce DCBH formation. This is not surprising as the SFR of a galaxy automatically folds in its recent SFH, and thus the shape of the SED and the resulting H$_2$-H$^-$ photodestruction rates. Galaxies with SFR $<0.1\msun \rm\ yr^{-1}$ would require the pristine atomic cooling halo to be present less than a kpc away from it. On the other hand, galaxies with SFR $>1\msun \rm\ yr^{-1}$ require the pristine atomic cooling halo to be located $\sim 3$~kpc away. Note that an equivalent relation between the stellar mass of a galaxy and D$_{max}$ can also be similarly derived, owing to the correlation between the stellar mass and star formation rate in form of the MS (Fig.~\ref{fig.MS}).

We also present a simple mathematical treatment for radiation emanating from multiple galaxies.  Using Eq.~\ref{eq.multiple} enables one to determine if a pristine atomic cooling halo in the vicinity of multiple galaxies can undergo DCBH formation, if the SFR of each galaxy and the distance of the pristine atomic cooling halo from each galaxy is known. The formula can be implemented in simulations, and can also be applied to observationally determine sites of DCBH formation in the vicinity of singular, or resolved objects such as CR7 where multiplicity is evident \citep{Sobral15a}.

\section*{Acknowledgements}
The authors are grateful to the referee for her/his very useful comments. BA would like to thank Mattis Magg, Anna Schauer, Kazu Omukai and Takashi Hosokawa for insightful discussion. BA and RSK acknowledge funding from the European Research Council under the European Community's Seventh Framework Programme (FP7/2007-2013) via the ERC Advanced Grant STARLIGHT (project number 339177).
\bibliographystyle{mn2e}
\bibliography{babib}

\begin{thebibliography}{36}
\expandafter\ifx\csname natexlab\endcsname\relax\def\natexlab#1{#1}\fi

\bibitem[{Agarwal {et~al}\mbox{.}(2017)Agarwal, Cullen, Khochfar, Klessen,
  Glover, \& Johnson}]{Agarwal_binary}
Agarwal B., Cullen F., Khochfar S., Klessen R.~S., Glover S. C.~O., Johnson J.,
  2017, MNRAS, 468, L82

\bibitem[{Agarwal {et~al}\mbox{.}(2014)Agarwal, Dalla~Vecchia, Johnson,
  Khochfar, \& Paardekooper}]{Agarwal14}
Agarwal B., Dalla~Vecchia C., Johnson J.~L., Khochfar S., Paardekooper J.-P.,
  2014, MNRAS, 443, 648

\bibitem[{Agarwal \& Khochfar(2015)}]{Agarwal15a}
Agarwal B., Khochfar S., 2015, MNRAS, 446, 160

\bibitem[{Agarwal {et~al}\mbox{.}(2012)Agarwal, Khochfar, Johnson, Neistein,
  Dalla~Vecchia, \& Livio}]{Agarwal12}
Agarwal B., Khochfar S., Johnson J.~L., Neistein E., Dalla~Vecchia C., Livio
  M., 2012, MNRAS, 425, 2854

\bibitem[{Agarwal {et~al}\mbox{.}(2016)Agarwal, Smith, Glover, Natarajan, \&
  Khochfar}]{Agarwal15b}
Agarwal B., Smith B., Glover S. C.~O., Natarajan P., Khochfar S., 2016, MNRAS,
  459, 4209

\bibitem[{Ceverino {et~al}\mbox{.}(2017)Ceverino, Glover, \&
  Klessen}]{2017MNRAS.470.2791C}
Ceverino D., Glover S. C.~O., Klessen R.~S., 2017, MNRAS, 470, 2791

\bibitem[{Ceverino {et~al}\mbox{.}(2018)Ceverino, Klessen, \&
  Glover}]{2018MNRAS.tmp.2027C}
Ceverino D., Klessen R.~S., Glover S. C.~O., 2018, MNRAS

\bibitem[{Ceverino {et~al}\mbox{.}(2019)Ceverino, Klessen, \&
  Glover}]{2019MNRAS.484.1366C}
Ceverino D., Klessen R.~S., Glover S. C.~O., 2019, MNRAS, 484, 1366

\bibitem[{Ceverino \& Klypin(2009)}]{2009ApJ...695..292C}
Ceverino D., Klypin A., 2009, ApJ, 695, 292

\bibitem[{Cullen {et~al}\mbox{.}(2017)Cullen, McLure, Khochfar, Dunlop, \&
  Dalla~Vecchia}]{2017MNRAS.470.3006C}
Cullen F., McLure R.~J., Khochfar S., Dunlop J.~S., Dalla~Vecchia C., 2017,
  MNRAS, 470, 3006

\bibitem[{Davis {et~al}\mbox{.}(2014)Davis, Khochfar, \& Dalla}]{Andrew14}
Davis A.~J., Khochfar S., Dalla V.~C., 2014, MNRAS, 443, 985

\bibitem[{Eldridge {et~al}\mbox{.}(2017)Eldridge, Stanway, Xiao, McClelland,
  Taylor, Ng, Greis, \& Bray}]{bpassv2.1}
Eldridge J.~J., Stanway E.~R., Xiao L., McClelland L. A.~S., Taylor G., Ng M.,
  Greis S. M.~L., Bray J.~C., 2017, PASA, 34, e058

\bibitem[{Elliott {et~al}\mbox{.}(2012)Elliott, Greiner, Khochfar, Schady,
  Johnson, \& Rau}]{Jonny12}
Elliott J., Greiner J., Khochfar S., Schady P., Johnson J.~L., Rau A., 2012,
  A{\&}A, 539, A113

\bibitem[{Ferland {et~al}\mbox{.}(2017)Ferland, Chatzikos, Guzm{\'a}n, Lykins,
  van Hoof, Williams, Abel, Badnell, Keenan, Porter, \& Stancil}]{ferland2017}
Ferland G.~J. {et~al.}, 2017, Revista Mexicana de Astronom{\'\i}a y
  Astrof{\'\i}sica Vol. 53, 53, 385

\bibitem[{Habouzit {et~al}\mbox{.}(2016)Habouzit, Volonteri, Latif, Dubois, \&
  Peirani}]{Habouzit16hSAM}
Habouzit M., Volonteri M., Latif M., Dubois Y., Peirani S., 2016, MNRAS, 463,
  529

\bibitem[{Haiman {et~al}\mbox{.}(2000)Haiman, Abel, \& Rees}]{Haiman:2000p87}
Haiman Z., Abel T., Rees M.~J., 2000, ApJ, 534, 11

\bibitem[{Johnson {et~al}\mbox{.}(2013)Johnson, Dalla, \&
  Khochfar}]{Johnson:2013p2049}
Johnson J.~L., Dalla V.~C., Khochfar S., 2013, MNRAS, 428, 1857

\bibitem[{Kravtsov(2003)}]{2003ApJ...590L...1K}
Kravtsov A.~V., 2003, ApJ, 590, L1

\bibitem[{Kravtsov {et~al}\mbox{.}(1997)Kravtsov, Klypin, \&
  Khokhlov}]{1997ApJS..111...73K}
Kravtsov A.~V., Klypin A.~A., Khokhlov A.~M., 1997, ApJS, 111, 73

\bibitem[{Latif {et~al}\mbox{.}(2014)Latif, Bovino, Van~Borm, Grassi,
  Schleicher, \& Spaans}]{2014MNRAS.443.1979L}
Latif M.~A., Bovino S., Van~Borm C., Grassi T., Schleicher D. R.~G., Spaans M.,
  2014, MNRAS, 443, 1979

\bibitem[{Latif {et~al}\mbox{.}(2013)Latif, Schleicher, Schmidt, \&
  Niemeyer}]{Latif:2013p2787}
Latif M.~A., Schleicher D. R.~G., Schmidt W., Niemeyer J., 2013, MNRAS, 433,
  1607

\bibitem[{Machacek {et~al}\mbox{.}(2001)Machacek, Bryan, \&
  Abel}]{Machacek:2001p150}
Machacek M.~E., Bryan G.~L., Abel T., 2001, ApJ, 548, 509

\bibitem[{Oh \& Haiman(2002)}]{Oh:2002p836}
Oh S.~P., Haiman Z., 2002, ApJ, 569, 558

\bibitem[{Omukai(2001)}]{Omukai:2001p128}
Omukai K., 2001, ApJ, 546, 635

\bibitem[{O'Shea \& Norman(2008)}]{OShea:2008p41}
O'Shea B.~W., Norman M.~L., 2008, ApJ, 673, 14

\bibitem[{Paardekooper {et~al}\mbox{.}(2015)Paardekooper, Khochfar, \&
  Dalla}]{JP2015}
Paardekooper J.-P., Khochfar S., Dalla V.~C., 2015, MNRAS, 451, 2544

\bibitem[{Paardekooper {et~al}\mbox{.}(2013)Paardekooper, Khochfar, \&
  Dalla~Vecchia}]{Paardekooper:2013p2455}
Paardekooper J.-P., Khochfar S., Dalla~Vecchia C., 2013, MNRAS, 429, L94

\bibitem[{Regan \& Haehnelt(2009)}]{Regan:2009p776}
Regan J.~A., Haehnelt M.~G., 2009, MNRAS, 396, 343

\bibitem[{Schaye {et~al}\mbox{.}(2010)Schaye, Dalla~Vecchia, Booth, Wiersma,
  Theuns, Haas, Bertone, Duffy, McCarthy, \& van~de Voort}]{Schaye:2010p2481}
Schaye J. {et~al.}, 2010, MNRAS, 402, 1536

\bibitem[{Shang {et~al}\mbox{.}(2010)Shang, Bryan, \& Haiman}]{Shang:2010p33}
Shang C., Bryan G.~L., Haiman Z., 2010, MNRAS, 402, 1249

\bibitem[{Sobral {et~al}\mbox{.}(2015)Sobral, Matthee, Darvish, Schaerer,
  Mobasher, R{\"o}ttgering, Santos, \& Hemmati}]{Sobral15a}
Sobral D., Matthee J., Darvish B., Schaerer D., Mobasher B., R{\"o}ttgering H.
  J.~A., Santos S., Hemmati S., 2015, ApJ, 808, 139

\bibitem[{Speagle {et~al}\mbox{.}(2014)Speagle, Steinhardt, Capak, \&
  Silverman}]{2014ApJS..214...15S}
Speagle J.~S., Steinhardt C.~L., Capak P.~L., Silverman J.~D., 2014, ApJS, 214,
  15

\bibitem[{Sugimura {et~al}\mbox{.}(2015)Sugimura, Coppola, Omukai, Galli, \&
  Palla}]{Sugimura:2015ut}
Sugimura K., Coppola C.~M., Omukai K., Galli D., Palla F., 2015, arXiv

\bibitem[{Sugimura {et~al}\mbox{.}(2014)Sugimura, Omukai, \&
  Inoue}]{Sugimura:2014p3946}
Sugimura K., Omukai K., Inoue A.~K., 2014, MNRAS, 445, 544

\bibitem[{Wolcott-Green {et~al}\mbox{.}(2017)Wolcott-Green, Haiman, \&
  Bryan}]{Wolcott2017}
Wolcott-Green J., Haiman Z., Bryan G.~L., 2017, MNRAS, 469, 3329

\bibitem[{Yoshida {et~al}\mbox{.}(2003)Yoshida, Abel, Hernquist, \&
  Sugiyama}]{Yoshida:2003p51}
Yoshida N., Abel T., Hernquist L., Sugiyama N., 2003, ApJ, 592, 645

\end{thebibliography}
\end{document}